# Digital Inheritance in Web3: A Case Study of Soulbound Tokens and the Social Recovery Pallet within the Polkadot and Kusama Ecosystems

**Justin Goldston**
National University

**Tomer Jordi Chaffer**
McGill University

**Justyna Osowska**
Women in Blockchain Canada

**Charles A. Von Goins II**
Rochester Institute of Technology

*In recent years, discussions centered around digital inheritance have increased among social media users and across blockchain ecosystems. As a result, digital assets such as social media content, cryptocurrencies, and non-fungible tokens have become increasingly valuable and widespread, leading to the need for clear and secure mechanisms for transferring these assets upon the testator's death or incapacitation. This paper presents an in-depth analysis of digital inheritance within the realm of Web3, focusing on the unique capacities of the Polkadot and Kusama ecosystems. By leveraging the native Social Recovery Pallet, the study highlights an innovative approach to safeguarding digital assets while pointing out possible exploitable weaknesses. The research underscores the limitations of traditional estate planning tools in digital asset management, calling for the formulation of updated, digital-focused estate documents. Several potential areas for future research are identified, including user and legal team education on digital inheritance, incorporation of Web2 assets, cross-blockchain interoperability, and the necessity for heightened security protocols. The paper concludes with a proposition for the development of a comprehensive "digital inheritance lego framework", integrating an array of blockchain-based solutions.*

**INTRODUCTION**

The notion of digital inheritance in the traditional Web2 world has gained increased prominence in recent years due to the rise of digital technology and the ubiquity of digital assets in the lives of individuals and families. In addition, with the growth of the Internet of Things (IoT), more and more devices are now storing personal data and assets, making it increasingly necessary to consider what will happen to these digital assets upon one's death. Digital assets refer to any property that exists in digital form. Examples of digital assets include online accounts (such as email, social media, and banking accounts), digital currency, non-fungible tokens (NFTs), and digital media such as photos, videos, and music (Al Harthy et al., 2022).

In the past, paper documents and tangible, physical assets were the primary legal instruments for an inheritance, and digital assets have often been excluded (Berlee, 2017). Additionally, laws highlighting digital assets upon death vary among states, countries, and even service providers such as Facebook, LinkedIn, YouTube, and Google. Conventional digital assets are often stored on third-party servers or in the cloud, making them sometimes difficult to track and transfer. Google, Apple, and Microsoft have all developed tools to allow users to designate a digital executor who can access accounts in the event of one's death (Brown, 2019).

As the term Web3 and decentralized digital assets such as cryptocurrencies and non-fungible tokens (NFTs) have begun to emerge in estate planning discussions, policymakers and practitioners are sometimes unclear on how to distribute these digital assets to a testator's heirs. In this study, using the Polkadot and Kusama Ecosystems as the foundation, we propose a digital inheritance plan that testators can use as a framework when working with the executors of their estate to manage digital assets. In the remainder of the section, we will discuss the motivation and potential contributions in conducting this research.

The remainder of the study is organized as follows: Section II introduces researchers to terms and topics that will be referred to throughout the study. Section III reviews the parties or entities involved in the digital inheritance process. Section IV reviews the self-sovereign identity (SSI) principles that serve as the foundation of the Web3 space. Section V expands on the SSI principles with Identity Schemes developed within the blockchain and Web3 space. Section VI introduces researchers to soulbound tokens (SBTs), while Section VII highlights the developments of social recovery wallets. Next, Section VIII takes a look at the current digital inheritance landscape across a number of blockchain ecosystems.

The focus of the study is included in Section XI, where the social recovery pallet within the Polkadot and Kusama ecosystems is introduced. Finally, Section X offers future research considerations based on the findings, and the study concludes with the acknowledgment and closing thoughts with Sections XI and XII.

**Motivation**

The primary motivation for writing on digital inheritance is to ensure that individuals, organizations, and families are aware of the importance of digital inheritance and that appropriate plans can be put in place for the smooth transfer of digital assets upon death or incapacitation. While the topic of digital inheritance in the area of social media has been studied in legal and technological contexts, little research has been performed in the areas of NFTs and cryptocurrencies (Saidakhrarovich, 2022). Digital inheritance is a growing concern for many individuals in the digital age as our lives become increasingly intertwined with digital technologies. This has resulted in a plethora of digital assets, such as online accounts, emails, and digital photos, which can be lost after a person's death.

In an often discussed topic regarding estate planning and digital inheritance, banking heir Matthew Mellon saw his US$2 million investment in Ripple (XRP) increase to approximately US$1 billion at the peak of the cryptocurrency market (Candan & Yaşlak, 2021). After unexpectedly passing away in 2019, Mellon's heirs would be in line to inherit this fortune in traditional law. The issue arose when it was uncovered that Mellon did not share his private keys with anyone and kept them on devices across the United States (Khanum & Mustafa, 2023). In an uncommon scenario, attorneys discovered that Ripple, the issuer of XRP, managed Mellon's assets due to Mellon's contribution to the project's early days. Although the attorneys could recover the assets, an agreement between Ripple and Mellon prevented the attorney from selling all of the assets at once, leading to the investment losing almost two-thirds of its value (Njui, 2018). In this case, if Mellon had a plan in place for his digital assets, the loss in the value of the assets may have been prevented.

Additionally, in a study by the Cremation Institute (2020), 89% of cryptocurrency investors worry about dying with their assets, while only 23% have a documented plan. Of that percentage, only 7% have created a will that includes cryptocurrencies. In other studies, roughly 60% of Americans and 50% of Canadians do not have a conventional will, which could be higher in other parts of the world (Lee et al., 2019; Wilson, 2022). With these findings, what are the reasons why so many individuals neglect to ensure that their assets are appropriately allocated to the rightful heirs and ensure their legacy is remembered?

In assessing why many individuals do not create a will could be due to the fear of death. In one study on SBTs, Chaffer and Goldston (2022) posited that the terror management theory (TMT) may alleviate the fear of death. TMT is a social and evolutionary psychology model that explains how people protect themselves against thoughts of death (Qi, 2023). With the evolution of NFTs within the Web3 space, individuals can develop a digital legacy with SBTs, and users may be able to create an immortal SSI for individuals with concerns about death.

**Contributions**

*Contributions to Individual Estate Planning*

Rapid global digitalization has altered how individuals communicate, do business, and store information. As a result, the digital world has spawned a new category of assets that should be accounted for in estate planning strategies. As more of an individual's life exists in digital form, the subject of how to handle digital assets has become crucial. Traditional estate planning executors typically account for the allocation of tangible property, but digital inheritance is sometimes overlooked.

Digital inheritance is a complex issue with far-reaching implications. Individuals and their advisors must understand the importance of digital inheritance and the need for updated estate planning physical and digital documents. By creating a plan for digital assets, individuals and their families can secure digital property transfer and protect their legacy in the digital age.

*Contributions to the Legal Practice*

The legal complexities of digital assets can make digital inheritance a daunting task. Estate planning documents should account for the various laws governing digital property. For example, the Stored Communications Act prohibits service providers from disclosing the contents of electronic communications without the account holder's consent (Korol, 2022). In addition, the laws governing digital assets vary from state to state. Some states within the United States have adopted the Uniform Electronic Wills Act, which allows individuals to create digital wills that are legally binding (Hirsch, 2020). Other states have adopted the Revised Uniform Fiduciary Access to Digital Assets Act, which allows fiduciaries to access and manage digital assets on behalf of an estate (Sheridan, 2020). With the differences in laws and regulations centered around digital inheritance, individuals and their advisors need to understand the applicable laws in their jurisdiction when creating an estate plan.

## PRELIMINARIES

### Blockchain

Blockchain is a distributed, immutable, and cryptographically secure digital ledger that records transactions across a network of computers in a secure and tamper-proof way (Rahardja et al., 2021). Through consensus protocols such as proof of work and proof of stake, nodes - or individuals - across blockchain networks validate and record transactions on the ledger, enabling individuals and businesses to streamline processes, increase transparency, and decrease costs.

### Polkadot

Polkadot is a blockchain-based protocol that enables the creation and interoperability of multiple blockchains. Known as a Layer Zero protocol, Polkadot's primary feature is the ability to enable cross-chain communication through cross-consensus and cross-chain messaging (XCM) between different blockchains (Tsepeleva & Korkhov, 2022). The Polkadot blockchain is made up of two main components: a Relay Chain and Parachains. The Relay Chain serves as the network's central validating chain, ensuring consensus and governance. Parachains are independent blockchains connected to the Relay Chain and can run applications written in the Polkadot runtime (Wood, 2016). The Polkadot protocol is intended to be a platform for Web 3.0 application development and provide an environment for developing secure, decentralized applications (dapps).

### Kusama

Kusama is an open-source, permissionless, proof-of-stake-based blockchain network. Kusama is modeled after the Polkadot blockchain and provides users with a platform for creating and deploying their dapps. Known as Polkadot's canary network, Kusama allows users to experiment with new technologies, such as governance and on chain upgrades, without the risk of major network disruptions (Teisserenc & Sepasgozar, 2022). Like Polkadot, the network offers scalability, security, and interoperability between blockchains (Anoop & Goldston, 2022). As a result, many projects deploy on Kusama, gain user feedback through usage, and make adjustments before migrating to the Polkadot blockchain network.

### Smart Contracts

Smart contracts can be referred to as agreements that can be automatically executed when certain criteria are satisfied on blockchain-based platforms. The provisions of the agreement cannot be amended or altered once they have been agreed upon since they are self-executing, secure, and unchangeable (Hewa et al., 2021). A range of transactions, including financial and legal agreements, can be facilitated and automated through the use of smart contracts.

### Soulbound Tokens

SBTs are cryptographic tokens permanently tied to a specific user or device. SBTs are typically used to represent ownership and are commonly associated with blockchain-based applications. SBTs are created using cryptographic algorithms, ensuring that no other user or device can access the token or the associated data (Weyl et al., 2022). SBTs are also resistant to replication, which further enhances the security of the token.

### Non-Fungible Tokens

NFTs are cryptographic tokens that are used to represent digital assets with unique characteristics that can be used to authenticate and track ownership and transfer of digital assets. They are distinct from fungible tokens, representing a common unit of value that can be exchanged on a blockchain. Unlike fungible tokens, each NFT is cryptographically unique and can be used to track and transfer ownership of digital assets (Valeonti et al., 2021). NFTs are held and validated on a blockchain and can represent collectibles, digital artwork, game items, real estate, or digital identities.

**Cryptocurrencies**

Used as a form of utility across a number of blockchain protocols, cryptocurrencies are digital assets that are created and exchanged using cryptography and distributed ledger technology. These decentralized digital assets have no connection to any central bank or government and rely on peer-to-peer networks for transaction transfer and validation (Panda et al., 2023). Cryptocurrencies are designed to be secure, with anonymous and irreversible transactions that eliminate the need for third-party intermediaries such as banks or other financial institutions. Although Central Bank Digital Currencies (CBDCs) in jurisdictions use stablecoins - a class of cryptocurrency - the reference to cryptocurrencies in this study will be focused on decentralized cryptocurrencies.

**Social Recovery**

The act of social recovery on blockchain protocols refers to the ability for users to recover assets from an account in the event that they lose their private keys or forget their password. Through cryptographic protection, social recovery wallets can provide a secure and efficient way to restore users' identities, access, and other important data (Ngo et al., 2023). Additionally, social recovery in blockchain can allow for the re-establishment of trust between users and their respective networks and provide a secure way for users to store and manage their personal data and assets.

**Self-Sovereign Identity**

SSI is a decentralized authentication system that enables individuals to securely, privately, and autonomously control their personal identity data. SSI is based on the principles of self-ownership, portability, and control, which means that individuals own and manage their information and can move it between service providers without requiring a centralized entity to act as an intermediary (Zhuang et al., 2023). By giving individuals more control over their personal data, SSI can be seen as a way to help individuals gain access to services, protect their privacy, and combat identity theft.

**Multi-Signature**

Multi-signature (multi-sig) is a cryptographic technique that requires multiple individuals to confirm a transaction through authorized signatures before the transaction is executed. Multi-sig is used across various blockchain-based networks to allow for a more decentralized governance model with fewer points of failure and is also used in various applications such as shared accounts, escrow services, and smart contracts (Kansal et al., 2022). The multi-sig standard follows the m-of-n ratio - also known as the quorum quotient. For example, a common multi-sig configuration is 3-of-5, which means three private keys - individuals - would have to sign a transaction before it is executed. If a transaction did not receive the required number of signatures from users holding private keys, the transaction would not execute.

**Seed Phrase**

A seed phrase, also known as a recovery phrase or backup phrase, is a string of words that can be used to regain access to a cryptocurrency wallet. Seed phrases are a method of backing up and securing the private keys that allow access to the wallet's funds, and the order of the words in the phrase must be entered correctly to regain access to the wallet (Shaik, 2020). A seed phrase is typically made up of 12 to 24 words and is generated when the wallet is first created.

**ENTITIES**

In traditional inheritance proceedings, several parties are involved in the digital inheritance planning process. Although many parties may remain the same, the entities involved in traditional and digital inheritance planning will be reviewed.

**The Testator**

The testator is the individual working with another party to develop a will. For a will to be legally valid, it must be executed by the testator in accordance with the laws of the jurisdiction where the will is being made (Smyth & Peiros, 2017). This typically involves the testator signing the will in the presence of witnesses, who must also sign the will to validate it. The testator can also appoint guardians for minor children and designate beneficiaries to receive specific assets or a share of the estate. The testator may also include provisions for charitable gifts, trusts, and other arrangements in the will.

The testator should be very explicit in his or her wishes regarding how they would like for digital assets to be transferred. Based on where the testator is located, there are a number of existing legal provisions governing the transfer of digital assets. For example, The General Data Protection Regulation (GDPR) is the primary legal instrument in Europe that provides individuals with the right to access their personal data and the right to have their data erased upon their death (Kharitonova, 2021). Although digital assets such as online accounts, emails, and digital photos have been identified in these legal instruments, researchers and legal practitioners have identified gaps within the existing frameworks when attempting to understand the legal provisions centered around digital assets such as cryptocurrencies and NFTs (Kurniawan et al., 2023). The legal frameworks regarding digital inheritance will be discussed in greater detail in subsequent sections of this study.

**The Executor**

An executor for an estate is a person responsible for carrying out the wishes of a deceased person as outlined in their will in accordance with the law. This includes gathering the deceased's assets, paying any debts, and distributing the remaining assets to the beneficiaries. The executor must also file any necessary paperwork with the court and handle any disputes that may arise.

For digital assets, this role will remain the same. Although an individual could work with their trustees - or guardians - to have a smart contract executed based on agreed-upon terms, it may be advisable to have a digital executor included in the process (Kurniawan et al., 2023). As laws change based on geographical region, legal guidelines give executors the right to access digital assets and require online service providers to provide access to digital assets upon request (Saidakhrarovich, 2022). They also provide clear guidance on managing digital assets in the event of a death.

**The Trustee**

The trustee is the individual or entity responsible for managing and administering the assets of a trust. The trustee may be appointed to manage the estate's assets on behalf of the beneficiaries. Working closely with the estate executor, the trustee's role is to act in the beneficiaries' best interests and ensure that the trust assets are managed and distributed according to the trust document's terms. There are several different types of trustees that may be appointed to manage an estate, including individual trustees, corporate trustees, and co-trustees (Johnson, 2023). The type of trustee best suited for a particular situation will depend on the size and complexity of the estate, as well as the needs and preferences of the beneficiaries.

Trustees are sometimes referred to as guardians for solutions that support digital inheritance. The difference between guardians compared to traditional trustees is that guardians can use a multi-sig approach to approve the release of assets upon the testator's death (Buterin, 2021b). Multi-sigs can provide an additional layer of privacy and security to both the testator and the guardians (Sun et al., 2022). Although the testator can identify the trustees or guardians during the digital inheritance planning process, per the wishes identified by the testator, guardians do not have to know the identities of the other guardians.

**The Beneficiaries**

The beneficiaries of an estate are the individuals who are legally entitled to the testator's assets. Family members are typically included, such as spouses, children, and other close relatives. Friends, charities, and other organizations could also be beneficiaries. The executor of the estate is responsible for ensuring that the assets are distributed in accordance with the deceased's wishes. Beneficiaries may be entitled to a lump sum payment or periodic distributions (Gross et al., 2017). Beneficiaries may receive a combination of both in some cases.

In regards to digital assets, researchers have been looking into the importance of SSI principles for testators further to increase the secure transfer of assets to their beneficiaries. SSI principles provide a framework for digital inheritance, allowing individuals to securely and safely transfer their digital assets without relying on third-party intermediaries (Soltani et al., 2021). SSI principles involve the use of secure identity layers, effective data privacy, and secure protocols built on distributed ledger technology (Satybaldy et al., 2020). These layers ensure that valuable assets are transferred to the verifiable digital identities of the testator's beneficiaries.

**SSI PRINCIPLES**

Shuaib et al. (2021) developed a framework of SSI principles that users should consider when sharing data on the internet. This framework could also be considered when joining a decentralized Web3 ecosystem and developing a digital inheritance strategy. Shuaib et al.'s framework consists of the following principles.

**Control**
   The SSI framework is based on the belief that individuals should have control over their digital identity and be able to have full control of their data and digital assets. Many studies have highlighted how centralized entities have changed terms of service policies that enable them to share a user's primary data and metadata (Acker, 2018; Weller & Kinder-Kurlanda, 2016). Guidi (2020) identified the potential risks of over three billion social media users regarding the control, security, and safety of their privacy and proposed a decentralized blockchain-based online social network to mitigate current risks. While this could potentially control conventional digital assets such as photos and videos, more valuable assets such as cryptocurrencies and NFTs may require even greater control.

**Access**
   Users should be able to access their data at any time easily. In the area of digital inheritance, guardians or trustees should also be able to access a testator's data or assets at any time as well, with a few caveats. To provide an additional layer of security, a testator can choose to delay access to digital assets by setting up an account that could be tied to their social recovery wallet (Buterin, 2021b). Based on the number of days determined by the testator, guardians could cancel the transaction if foul play or a potential hack is of concern.

**Transparency**
   Transparency is essential for the proper functioning of SSI as it enables individuals to make informed decisions about their personal data and identity information. Without transparency, individuals may not be aware of how their data is being used or shared, and may not have the ability to control or protect their data. One of the main challenges with the transparency of digital assets is the visibility of the assets in one place. As users will have social media passwords, data, and content across various platforms, along with cryptocurrencies and NFTs across a number of different blockchains, it could be difficult to aggregate all of this content into one application. As projects have created dashboards for tracking decentralized finance (DeFi) portfolios across a number of different blockchain networks, the potential to create digital inheritance dashboards to increase the transparency of a user's digital assets will be reviewed in further detail within the Discussion section of this study.

**Persistence**
   Persistence is a concept based on the idea that an individual's identity is maintained throughout their lifetime, regardless of changes in personal circumstances. Persistence is a critical element of SSI, as it ensures that the individual's identity is consistently linked to their digital assets, even after death. This ensures that the digital assets can be securely transferred to the designated beneficiaries. Researchers have also mentioned the use of biometric authentication and blockchain technology to ensure that the individual's identity is securely linked to their digital assets (Lee & Jeong, 2021). When planning for digital inheritance, an individual could set up biometric authentication for themselves and the executor of their estate to ensure persistence and an additional layer of security for their assets. Given that some studies are theoretical, it is important to consider the legal implications of persistence in the context of digital inheritance and succession planning within one's geographical region.

**Portability**
   SSI requires that individuals can securely transfer their digital assets to other individuals or entities, as well as the ability to revoke access to their assets. Portability could become complicated for Web2 digital assets are stored in different services and platforms such as YouTube, TikTok, and Instagram. For heirs to access these assets, they must be able to access the same services and platforms that the deceased used. While storing one's digital identities and passwords across different services and platforms, this ensures that heirs will be able to access the digital assets of the deceased.

**Interoperability**
   Interoperability for SSI will be essential as the decentralized Web3 space continues to expand. Currently, a small number of blockchain networks provide cross-chain communication and the bridging of assets (Chervinski et al., 2021). Layer zero blockchains such as Polkadot, Kusama, and Cosmos have built blockchain networks with cross-chain communication in mind. This communication and transfer of funds primarily occur across blockchains within the network. Although bridges are being developed within these networks to communicate with blockchains such as the Bitcoin network, the Ethereum network, and others such as Avalanche, solutions to aggregate cross-chain assets may need to be considered for solutions such as those being developed for digital inheritance.

**Consent**
The principle of consent is based on the idea that individuals should control how their data is used and who can access it. It requires that individuals give explicit consent to any digital data associated with them and have the right to revoke this consent at any time during their life. Although researchers have noted that the scope and coverage are extremely limited (Emmert, 2022), regulations have continued to emerge in many countries as policymakers have seen the increased use and monetization of digital assets.

In the United States, the Revised Uniform Fiduciary Access to Digital Assets Act (RUFADAA) provides a framework for managing digital accounts in the event of death or disability. The act gives executors and trustees access to emails, chats, and direct messages *only* if the individual explicitly consented (Sehati, 2021; Sheridan, 2020). Additionally, under the RUFADAA, centralized entities such as Google and Meta are referred to as custodians of a user's data and may only provide access to the reasonably necessary information to settle an individual's estate and cannot provide access to deleted assets.

**Existence**
Proof of existence is a topic often discussed within the area of social media. As automated accounts or social media bots plague a number of platforms, bots could also pose a potential threat to SSI. Furthermore, as bots have been responsible for attacks on social media platforms, influencing public opinion, and potentially impacting elections through the spread of misinformation (Ali Alhosseini et al., 2019; Orabi et al., 2020), proof of existence for all parties included in a digital inheritance plan will be important.

One way users can demonstrate proof of existence is with SBTs. For example, in reviewing the use cases for SBTs, if the testator, executors, guardians, and beneficiaries had SBTs created by the testator and transferred to the users' wallets, that would be a part of the digital inheritance plan. This process could prove the existence of all users and increase the security of the testator's digital assets.

**Minimalization**
Minimalization refers to the idea that individuals should only be required to share the minimum amount of personal data necessary to accomplish a specific task or goal. In the context of SSI, minimalization means that individuals should be able to share only the necessary pieces of information about themselves in order to prove their identity or access certain services rather than having to disclose large amounts of personal data.

Within decentralized environments, zero-knowledge proofs enable ZKPs users to share a limited amount of information to prove they hold certain information or even one's identity. ZKPs are a type of cryptographic protocol that enables one person - the prover - to demonstrate to another - the verifier - that they possess specific knowledge or information without disclosing that knowledge or information (Alikhani et al., 2021). ZKPs may become vital for the digital inheritance of testators as they may not want to show all of their digital assets to beneficiaries or guardians but only want to prove that the contents within the account or wallet exist.

**Protection**
Numerous studies centered around social engineering attacks on social media platforms have been performed across various disciplines, such as Business, Marketing, and Computer Science. As the intention of these attacks, in many cases, is for monetary gain, researchers and practitioners have called for additional protections among social media users (Bushmelev et al., 2022; Zulkiffli et al., 2020). To counter these risks, legal and regulatory frameworks such as the General Data Protection Regulation (GDPR) in the European Union have been developed to protect individuals' data, such as social media accounts, online banking, and cloud storage (Diamantopoulou et al., 2018). However, while dapps can provide anonymity and protection by allowing a user to connect a wallet (Guidi, 2020), apps in blockchain networks such as Ethereum, Solana, Polygon, and Polkadot have yet to gain wide adoption among users.

For decentralized digital assets such as cryptocurrencies and NFTs, SafeHaven is a project within the VeChain blockchain ecosystem that has added an extra layer of protection for testators by implementing a deadman's switch (Digital Assets Daily, 2022). A deadman's switch keeps assets in a wallet or account safe by allowing the testator to define a period for the application to alert the user to enter authentication information and confirm liveness periodically. If the user fails to enter the authentication information within a certain period, the assets are automatically transferred to the designated beneficiaries.

## IDENTITY SCHEMES

To expand on Shuaib et al.'s (2021) SSI principles, Sun et al. (2022) identified the following identity schemes users should be aware of as Web3 and the decentralized web continue to evolve.

**Personality Identity**
Personality Identity is a digital identity model that gives individuals full control over their identity and related information. Personality identity allows users to manage their identity credentials and access their data online without relying on a third-party identity provider, such as a government or corporate entity. This allows users to fully monetize off of their personality identity, compared to the conventional model where corporate entities receive a large percentage of the monetization rewards (Kopf, 2020). With this approach, an identity management system of this type provides users with a high level of privacy and security by allowing them full control over their data and the services that use it.

**Credential Identity**
Credential identity provides a secure way to prove one's identity to other parties online. It is based on the idea that a digital identity is based on an individual's unique attributes, which are securely stored and managed by the individual, and is verifiable by other parties. Therefore, credential Identity allows individuals to create and store digital credentials that are cryptographically secured and can be used to prove qualifications and access services without having to rely on a third party. For example, services such as the POAP and Ethereum Name Service (ENS) have created solutions that prove attendance or participation at events or prove digital assets or credentials are held in a digital wallet the address - in an ENS case - is attached to (Béres et al., 2021; van Rijmenam, 2022).

**Reputational Identity**
Reputational Identity for SSI refers to the digital representation of an individual's identity that is built upon their past interactions, digital and physical, and is used to assess an individual's trustworthiness. Also referred to as proof of authority (PoA), blockchain protocols such as VeChain have developed consensus mechanisms around the PoA model (Manolache et al., 2022). When using the PoA consensus mechanism, nodes, or individuals, must have a reputational identity established and recognized by the network. This reputational identity serves as a means of validating the node's trustworthiness and is a key factor in determining the authority of the node.

**Data Identity**
Data Identity is a concept in which individuals control their digital identity data, including authentication, access management, and privacy. Projects such as Kilt Protocol provide decentralized identity (DID) tools for users within the Polkadot and Kusama ecosystems (Teisserenc & Sepasgozar, 2022; Thomas et al., 2022). This data is collected and stored in a secure, decentralized digital identity store. This store is used to verify the identity of users and enable other digital activities, such as secure transactions and digital signing.

**A NEW CASE FOR NON-FUNGIBLE TOKENS**

Non-fungible tokens (NFTs) are digital assets that can be used to represent ownership of real or virtual items. NFTs are unique, meaning each token is distinct and cannot be replaced with another. This makes NFTs different from traditional digital assets such as Bitcoin, Ethereum, and Polkadot tokens, which are interchangeable and divisible. NFTs became popular in 2021 and 2022 due to their ability to enable the monetization of digital items such as art, music, and gaming assets (Kharitonova, 2021). With blockchain infrastructure technology underlying it, an NFT guarantees a secure and immutable record of ownership for the asset being represented. NFTs can be viewed as a use case for blockchain that has enabled digital assets to be monetized and traded securely and transparently, paving the way for a new form of digital asset ownership.

As cryptocurrencies, decentralized finance, and NFTs have allowed users to amass wealth due to the ability to transfer digital assets, there may also be a use case for non-transferrable assets within the Web3 space. For example, within the Web2 gaming space, namely in World of Warcraft (WoW), soulbound items can be collected in-game but cannot be transferred to another player. Soulbound items are valuable because they give players access to exclusive items and provide additional benefits, such as increased experience or bonuses, when used in certain activities. However, with WoW continuing to bring in over a million daily active users (DAUs) in 2022 (Flunger et al., 2022), how can the concept of soulbound items transfer over to the Web3 world?

In an article called Soulbound, Buterin (2021a) introduced the concept of SBTs for governance mechanisms across blockchain protocols. Current state, because users can go onto centralized exchanges (CEXs) or decentralized exchanges (DEXs) to purchase governance tokens, parties that are driven by purposes potentially detrimental to the protocol could concentrate interests away from the ecosystem's community, defeating the purpose of the decentralized nature of blockchain and Web3. Furthermore, with the introduction of SBTs, similar to soulbound items in Web2 gaming, SBTs would be non-transferable. While it can be argued that a user could sell an entire wallet to another user, making governance tokens as SBTs may be a topic for further research.

With governance tokens providing one use case for SBTs, another use case for non-cryptocurrency users may be for proof of identity. Two projects using a de-facto soulbound framework within the Ethereum ecosystem are BrightID and Proof of Humanity (Sun et al., 2022). To prove that the owners of accounts are indeed actual individuals, in an SBT model, the token identifiers are not transferrable, given that on-chain assets are soulbound to each verified individual (Buterin, 2021a). As Web3 looks to introduce SSI to the masses, these tools' virtual asset service providers (VASPs) may consider an SBT framework.

As previously mentioned, SSI is a foundational tenant of Web3 that allows users to have complete control over their digital identities. By leveraging blockchain technology, SSI can provide access to services and applications in a more efficient manner, as well as enable improved trust between people, organizations, and governments around the world (Ferdous et al., 2019). Similar to how NFTs onboard new users into the blockchain and Web3 spaces, SSI with SBTs could encourage more users to enter various blockchain ecosystems.

**Additional Use Cases for Soulbound Tokens**

Upon the release of Buterin's article Soulbound, projects across Ethereum and other blockchain protocols began to develop solutions for SBTs. Within the United States, Wyoming recognized decentralized autonomous organizations (DAOs) as separate legal entities through a supplement to the law governing limited liability companies (LLCs) in 2021 (Chun et al., 2022). This led to a number of projects registering as DAOs in 2021 and 2022. DAOs such as Reputation DAO leveraged SBTs as reputation systems for utilities such as credit scores, proof of identity in the play-and-earn space, and proof of identity for council members across various ecosystems (Sun et al., 2022). In taking this approach with SBTs, blockchain can potentially provide further transparency to the on-chain behavior of users.

In another example of how SBTs can cross over into the Web3 space, RMRK, a project within the Polkadot and Kusama ecosystems, created an SBT called Soulbound 2.0 (SBT2). Developed with the metaverse and GameFi in mind, RMRK is another project that created a reputation-based SBT (Oli, 2022). As RMRK's SBT2 will gain attributes based on how long a user plays a blockchain-based game or stays in a metaverse, SBT2s can be viewed as dynamic SBTs. Phala Network, another project within the Polkadot and Kusama ecosystems, leveraged RMRK's SBT2 to create an SBT within their PhalaWorld metaverse. Using a Spirit, any user can obtain an SBT that can only be leveled up by performing activities throughout the Phala ecosystem. By demonstrating behaviors such as participating in on-chain activities, being a community advocate on Web2 social media platforms, contributing to the community from a technical perspective, or staking PHA - the native token for the Phala ecosystem, SBTs can only evolve with user interaction.

Within the Binance ecosystem, the Binance Account Bound (BAB) SBT was created for Binance users that completed their know your customer (KYC) verification (Sun et al., 2022). Given that Binance is a CEX that holds a number of assets across a number of wallets, could SBTs provide a use case for know-your-exchange (KYX) verification? With the collapse of FTX in 2022, users could track the suspicious activity of a wallet holding large amounts of FTT tokens - the native token of the FTX exchange. If CEXs such as Binance leverage the BAB SBT they created to provide increased transparency, other CEXs, DAOs, and protocols may have to follow similar processes to develop standards across the Web3 landscape.

**A Decentralized Society with Soulbound Tokens**

In considering use cases for a number of industries, Weyl et al. (Weyl et al., 2022) proposed that SBTs could potentially create the foundation for a decentralized society (DeSoc). Referred to as an extended resume, if users leveraged wallets as SBTs, these tools could provide the credentials and commitments to conduct a variety of activities in a decentralized fashion. Although decentralized finance activities are currently performed across a number of blockchain ecosystems, methods to track a user's credit score or repayment history do not currently exist. SBTs could provide a way to unlock lending opportunities for individuals and organizations that do not have the financial history or assets traditional financial institutions recognize. For example, if a user owns cryptocurrencies or NFTs, if they hold those assets in an SBT-based wallet, with the user's permission, the wallet could show the payment history and the 'crypto credit score' to the lender.

Tying SBTs back to extended resumes, college degrees, certifications, and transcripts could also be on-chain as SBTs. Given that degrees, certifications, and transcripts are soulbound in the Web2 world, employers and academic institutions could leverage SBTs to confirm that the contents of an applicant's resume are valid. Additionally, if an employer would want to conduct a reference check, the reference's addresses could be included in the SBT, where attestations could be performed on-chain.

As discussed in Section IV's Existence and Minimization areas, SBTs will play a key role in digital inheritance planning. While SBTs can be created for all entities that are a part of the executor's will, ZKPs may also be used in conjunction with SBTs to ensure that no one can manipulate the system (Xue & Wang, 2022). In addition, as SBTs will become another foundational piece in the digital inheritance framework, protocols and networks outside of those previously mentioned may consider SBTs to provide privacy and security to their users as they build out their digital inheritance plans.

**SOCIAL RECOVERY WALLETS AND ACCOUNTS**

In an effort to work toward SSI in the digital age, intuitive solutions must be created for non-technical users. Given that the Web3 future will be built on dapps, if a user loses their password or seed phrase, no intermediary or centralized entity can be contacted to recover a user's password. Projects have created social recovery wallets to mitigate the risk of lost assets. Social recovery wallets are smart contract wallets where a user's assets will be within that smart contract (Buterin, 2021b). A signing key protects access to the smart contract, and that signing key is used to approve transactions or the transfer of assets. If a user forgets his or her password, trustees - or guardians in this case - can come together to recover the account.

Within the Ethereum ecosystem, Argent and Loopring have developed multi-sig social recovery wallet solutions where a user can set up between three to seven guardians they can reach out to in order to recover an account (Buterin, 2021b). As new users entering the Web3 space may forget his or her wallet password or seed phrase, this function may be a helpful tool. Additionally, as passwords, pictures, video files, and other Web2-native digital assets could be stored in a social recovery wallet through the use of decentralized storage applications such as Arweave, Filecoin, or The InterPlanetary File System (IPFS), social recovery wallets may provide a more secure solution for assets during one's life. Building upon the utility of social recovery wallets, a use case for social recovery wallets could be explored for users performing digital inheritance planning activities.

**THE EXISTING DIGITAL INHERITANCE LANDSCAPE**

With more individuals understanding the importance of securing their digital assets, more projects and solutions have entered the market. Organizations have created solutions to help individuals create digital estate plans and transfer digital assets after death. Similar to traditional executors of a will or estate, a digital executor is responsible for an individual's digital assets (Browne & Darcy, 2020). The digital executor of the estate would identify and catalog digital assets, work with the individual to create a digital will, and manage the transfer of digital property (Zheng, 2023). This ensures that digital assets remain safe and accessible even after the account holder has passed away.

As researchers have analyzed projects that have offered digital inheritance solutions in the past, such as PassOn, DigiPulse, and TrustVerse, based on the research, only two digital inheritance solutions exist - Vault12 and SafeHaven (Singh et al., 2022).

**Vault 12**
Vault12 is a digital inheritance platform that enables users to securely store and transfer digital assets, such as cryptocurrency and NFTs, to a designated beneficiary. Vault12's solution also allows users to store digital information, such as passwords, bank accounts, documents, and photos (Skibinsky et al., 2018). In addition, by using a combination of multi-signature technology, Vault12 ensures that assets are securely transferred to a designated beneficiary upon the user's death or incapacitation.

**SafeHaven**
SafeHaven is a platform that provides individuals with an innovative way to protect and manage their digital assets in the event of death, disability, or other life-changing events. SafeHaven has developed a solution called Inheriti that enables users to securely store, share, and transfer their digital assets to their heirs. Schouppe (2022) patented the AES 256 cryptographic key encryption SafeHaven's solutions are built upon and leverage the VeChain blockchain Proof of Authority consensus mechanism.

In expanding outside of the VeChain ecosystem, SafeHaven looks to introduce its solutions to users across Ethereum Virtual Machine (EVM)-compatible blockchains (Digital Assets Daily, 2022). As projects outside the Ethereum ecosystem, such as Moonbeam within the Polkadot ecosystem, are building EVM solutions for non-native EVM blockchains, this approach may provide more users with digital inheritance solutions (Jia & Yin, 2022). Furthermore, given the ultimate goal of interoperability for various blockchain projects, using EVMs to bridge solutions such as Inheriti into different blockchain ecosystems may increase the awareness of digital inheritance offerings.

**CASE STUDY**

In order to demonstrate a proposed solution for a digital inheritance framework, a case study methodology was selected. In analyzing different methodologies, one reason for proposing a case study for using the social recovery pallet for digital inheritance is that it allows for exploring the unique characteristics and potential challenges of the specific context in which the pallet will be used. Also, a case study can be useful for generating recommendations and best practices for using the social recovery pallet in other contexts, as developers and users can be interviewed during the data collection and exploration process (Shrestha & Bhattarai, 2021). Using a case study approach and carefully examining the functionality of the social recovery pallet, the researchers hope that the proposed solution can lead to future implementations within and outside the Polkadot and Kusama ecosystems.

**Substrate**
In constructing the digital inheritance lego within the Polkadot and Kusama ecosystems, the proposed solution will be built upon the Substrate framework. The Substrate framework is a modular platform that allows developers to build custom blockchain applications similar to templates in other technological solutions (Wood, 2022). Pallets are self-contained units of code that provide specific functionality to the blockchain, such as governance, staking, or consensus (Johnson et al., 2019). Developers can choose from a wide range of pre-built pallets or create their own custom pallets to add specific features to their blockchain (Abbas et al., 2022). This modular design enables developers to easily customize their blockchains and add or remove features as needed.

In the Polkadot and Kusama ecosystems, pallets are implemented as WebAssembly (Wasm) modules that are compiled and uploaded onto the blockchain (Wood, 2016). The Substrate pallets are compiled Wasm code, which can be executed by the network runtime.

Pallets can be grouped into three categories:

1) Core pallets: Core pallets are the fundamental building blocks of the network, providing the necessary
2) infrastructure for consensus, networking, and other critical functions.
3) Runtime pallets: Runtime pallets are the pallets that provide the core functionality of the network, such as smart contract execution, governance, and staking.
4) Application pallets: Application pallets are the pallets that enable the development of specific dApps or use cases, such as financial services, supply chain management, or prediction markets (Burdges et al., 2020).

Substrate pallets have a wide range of potential uses and benefits in the context of the Polkadot and Kusama networks. For example, pallets have previously been created to develop decentralized governance models and create decentralized identity systems (Scott et al., 2023). In the next section of this case study, we will explore an additional use case for the social recovery pallet to develop a digital inheritance plan for users within the Polkadot and Kusama ecosystems.

**The Social Recovery Pallet**
Natively, similar to social recovery wallets previously mentioned, the social recovery pallet enables users on Substrate blockchains to be able to recover accounts in the event that he or she loses a password, private key, and their seed phrase. As decentralized blockchain protocols should not be able to retrieve an individual's private key, the social recovery pallet creates a new key and makes calls on behalf of one's last account to a new account for one's assets to be retrieved (Tabrizi, 2020). Fig. 1 provides an overview of the social recovery process.

The process includes an individual's existing account, where he or she will set up a *createRecovery* process that uses an M-of-N recovery tool. M-of-N is a multi-sig process where the user will identify a defined number of trustees - referred to as friends in the case of the social recovery pallet - that need to approve the recovery process to recover the account (Tabrizi, 2020). In developing a new use case for the social recovery pallet, the process is viewed through a digital inheritance lens moving forward.

# FIGURE 1
## AN OVERVIEW OF THE PUBKEY CHANGE PROCESS

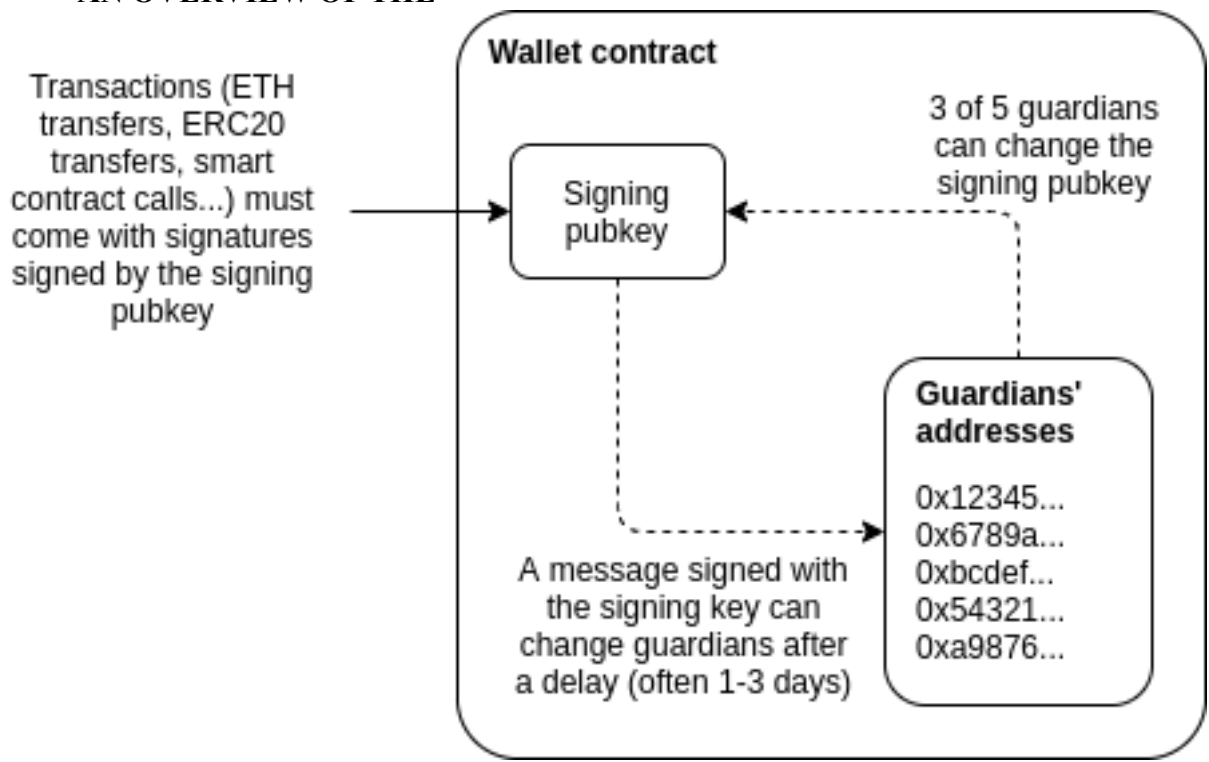

**Digital Inheritance Planning**

The first step in developing a digital inheritance strategy would be to set up a recovery configuration. In the configuration, a testator may work with a digital executor - who may also be set up as a friend - to identify the number of friends that would be added to the M-of-N tool (Lucsok, 2020). The 'M' would be the number of friends or beneficiaries the testator trusts, whereas the 'N' would be the threshold of users that must confirm the testator's death. For example, if the testator identified five friends he or she trusted to be gatekeepers of the account, three of them would have to approve the transaction for the delay period to begin.

For the native social recovery pallet, the holder sets the delay period to identify how long one would have to wait until the account could be recovered. In the case of setting up a digital inheritance plan, the testator could work with the digital executor to identify how long the beneficiaries would have to wait until the digital assets are transferred (Petrowski & Tabrizi, 2020). The delay period also provides an additional layer of security against social engineering attacks as the testator could identify malicious actors attempting to access his or her account during life. In this use case, given that the testator may be holding large amounts of assets in the account to be recovered, they may work with their digital executor to set a higher delay period.

Once the testator has set up the recovery configuration, they will be required to deposit KSM on the Kusama blockchain or DOT on the Polkadot blockchain to put the data on-chain. Additionally, upon death, the digital executor or a trustee - or friend - would have to issue a deposit to initiate the social recovery process. The deposit to initiate the social recovery process could also be deemed a security measure or a honeypot as if a malicious actor submits a deposit to begin the social recovery process, the testator or the digital executor could identify this attack and call the *closeRecovery* function to stop the process and receive the deposit.

**Executive the Digital Inheritance Process on Polkadot and Kusama**

Upon the testator's death, the digital executor could submit a deposit to initiate the social recovery process by calling the *initateRecovery* option. The digital executor could then contact all friends linked to the social recovery account to submit a *vouchRecovery* transaction [80]. Once the threshold set up during the M-of-N process has been reached and the delay period has passed, the digital executor would be able to access the digital assets from the old account and distribute the assets per the wishes of the testator through the *claimRecovery* process.

Once assets have been recovered from the old account, the *closeRecovery* process can be called, allowing the digital executor to recover the deposit needed to initiate the social recovery process. After all assets have been moved to the new account, the digital executor can call the *removeRecovery* option to remove the configuration and connection between the new account and the old account and make any further transactions or assets unrecoverable.

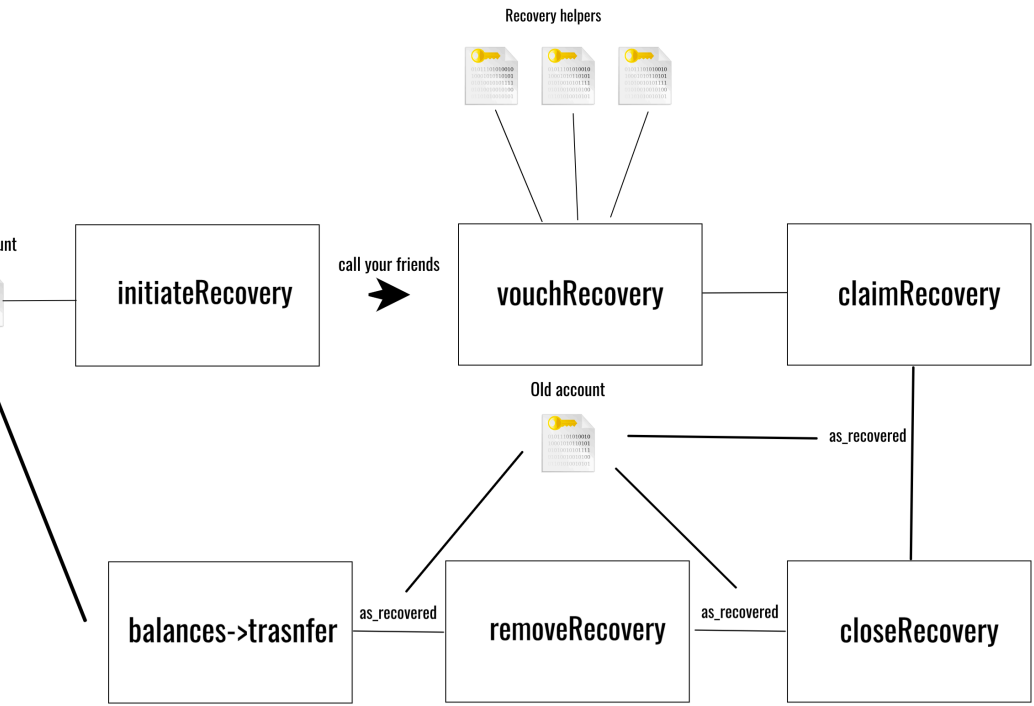

### FIGURE 2
### THE SOCIAL RECOVERY PROCESS ON POLKADOT AND KUSAMA

Although this is the proposed solution for setting up a digital inheritance plan, if a testator did not perform the recovery configuration, the developers of the social recovery pallet included an additional failsafe. In the event that the digital executor does not receive enough signatures on the *vouchRecovery* transaction from friends of the testator, the digital executor could go through the Polkadot or Kusama council or initiate a public proposal to use the Root origin to access the digital assets of the testator (Petrowski & Tabrizi, 2020). As this process would provide additional levels of complexity and the risk of attracting malicious actors, the proposed solution to develop a digital inheritance plan could provide the security and safety of digital assets that testators may expect.

**Limitations to the Native Social Recovery Pallet for Digital Inheritance**

While transparency across blockchain networks provides benefits in some use cases, it could attract unwanted attention in the case of digital inheritance. As the addresses of friends are currently public for the social recovery pallet, developers could modify the existing pallet to create an obfuscation layer to provide additional security. Given that users could create decentralized identities (DIDs) within the Polkadot and Kusama ecosystems using services offered by projects such as Latently, Kilt, and Ontology (Teisserenc & Sepasgozar, 2022; Thomas et al., 2022), digital executors and friends could be easily identifiable. In addition to external threats from malicious actors, if a testator did not want the selected friends to know others' identities, the current architecture of the social recovery pallet could allow friends to collaborate to initiate the social recovery process without the testator's or the digital executor's permission.

In the event that a testator's friends were to collude against the testator, if the testator did not periodically check to see if the social recovery process was initiated, current state, he or she would be unaware of any malicious acts (Tabrizi, 2020). Suppose a notification system was created to send alerts to the testator and the digital executor with preset reminders until on-chain action is taken. In that case, this addition could mitigate the risk of collusion.

Finally, the digital assets are not automatically transferred once a digital executor has accessed the old account and linked it to the new account. Currently, the new account is simply performing a call to the old account, and once all assets are removed from the old account, the *removeRecovery* function is called to remove the link between the two accounts. If a function was created within the social recovery pallet to transfer all assets using one function, this function might increase the intuitiveness of the pallet.

**The Determination and Implementation of the Deceased's Wishes (The Digital Will)**
Traditional estate planning instruments - such as wills and trusts - are not designed to address the transfer of digital assets. These instruments may provide instructions for the transfer of physical property, but they need to account for the complexities of digital property (Haley, 2023). To ensure that digital assets are effectively transferred, individuals must create updated estate planning documents that include specific instructions for managing digital property (Tatar & Akmaljon, 2023). These instruments should identify the types of digital assets owned by the individual, provide instructions for the transfer of digital assets, and name a digital executor to manage the digital estate (Komarevceva et al., 2022).

## FUTURE RESEARCH CONSIDERATIONS

Despite the proposed solution to leverage the social recovery pallet within the Polkadot and Kusama ecosystems for digital inheritance planning, effective ways to educate users and legal teams on this use case may require further research. Additionally, with the advancements in other technologies, such as artificial intelligence and quantum computing, the following topics may also require further investigation:

**Inclusion of Additional Digital Assets**
While Web3 native assets such as cryptocurrencies and NFTs can be easily stored in decentralized accounts and vaults, there is still a gap in the research as to how Web2 native assets such as social media accounts, content, and passwords can be stored. Phala Network within the Polkadot and Kusama ecosystems has developed external storage services using centralized tools such as Amazon S3 and decentralized tools such as Storj, Arweave, and Filecoin (Waller, 2022). However, additional research may be required regarding the security concerns and accessibility of storing different types of assets in a digital inheritance account and creating a type of pointer to those assets.

**Interoperability**
Interoperability is a common topic of discussion regarding Web3 and blockchain, but how will tools created for active accounts impact digital inheritance accounts? Additional research may be required to assess the feasibility of an account aggregator injecting digital assets from other blockchains once a social recovery process is initiated. For example, if a digital executor initiated the social recovery process, given the potential interoperability properties of the Polkadot and Kusama ecosystems, is it feasible for a social recovery pallet to perform calls to pre-defined wallet addresses on the Bitcoin, Ethereum, Avalanche, and Binance blockchains to transfer assets the social recovery account set up by the testator?

**Onboarding Attorneys to Establish Legal Frameworks**
As technology has grown, so have the legal issues surrounding digital inheritance. As a result, attorneys and digital executors can play a crucial role in helping individuals and families establish a digital inheritance strategy. They can also assist in developing more explicit and more defined legal frameworks centered around digital assets.

**Increased Security for Social Recovery Accounts and Vaults**
As awareness of users leveraging social recovery accounts for digital inheritance continues to grow, increased security measures may need to be implemented to prevent threats that may emerge in the future. To validate this point, The U.S. Secretary of Homeland Security, Alejandro Mayorkas, stated that leaders of organizations should be preparing now for the future threat of quantum computers (Clinton, 2023). While the National Institute of Standards and Technology (NIST) is currently developing post-quantum cryptography standards, the U.S. Cybersecurity and Infrastructure Security Agency (CISA) has developed a roadmap that leaders of organizations should follow in the near term (Barker et al., 2021). However, as there is currently no consensus as to whether quantum computing may be a threat to blockchains soon, further research may need to be performed on the topic.

Additionally, as organizations such as Apple have been using secure enclaves in the organization's mobile solutions, additional research may need to be conducted to understand whether additional layers of security should be introduced. Also referred to as trusted execution environments (TEEs), secure enclaves can be used to store cryptographic keys, protect sensitive data from unauthorized access, and enable secure remote attestation (Usman et al., 2023). With biometric authentication providing an additional security layer during one's life, secure enclaves may need to be considered to counter future threats once someone passes away.

**A Digital Inheritance Lego Framework**

Similar to how DeFi legos were built from existing protocols and dapps to create new protocols and solutions, additional research may be required to integrate a number of tools discussed in this study, such as SBTs, social recovery wallets, vaults, secure enclaves, Zk-proofs, and smart contracts. In addition, given these solutions exist across a number of blockchain ecosystems, composability among these applications may also need to be considered.

**ACKNOWLEDGEMENT**

The authors would like to thank Joshua Waller, Ray Lu, and other team members from Phala and Bit.Country from the Polkadot and Kusama ecosystems for their thoughts and input into this study. The authors would also like to thank Charlotte McCurdy of Perley-Robertson, Hill & McDougall for providing input from a legal aspect on digital asset inheritance. Finally, the authors would like to thank  Daniel Krupka of The Coin Bureau for insights from a user experience perspective. This work was supported by a research grant from the Web3 Foundation.

**CONCLUSION**

The findings from this study have important implications for individuals and those involved in managing digital assets after one's death. First, the findings suggest that individuals are beginning to understand the importance of adding Web2-native digital assets into their inheritance plans, but further education must take place for individuals, digital executors, and policymakers regarding Web3-native digital assets. Second, the findings suggest that although the study focused on the Polkadot and Kusama ecosystems, digital inheritance lego frameworks and SBT structures may be considered in other protocols to enable the composability and interoperability of integrating Web2 and Web3 digital assets using some aggregation approach. Finally, the study highlights the importance of understanding the motivations for digital inheritance. This understanding can inform effective strategies for managing digital assets after a person's death as the decentralized Web3 environment continues to evolve.

**REFERENCES**

Abbas, H., Caprolu, M., & Di Pietro, R. (2022, August). Analysis of Polkadot: Architecture, internals, and contradictions. In *2022 IEEE International Conference on Blockchain* (Blockchain), 61-70.

Acker, A. (2018). Data craft: The manipulation of social media metadata. *D&S Res. Inst., 13*(1), 1-26.

Al Harthy, K., Agarwal, A., Naidu, V. R., & Al Shuhaimi, F. (2022, October). Trading digital assets owned by Omani entrepreneurs using non-fungible tokens (NFT): A proposed framework. In *2022 10th International Conference on Reliability*, Infocom Technologies and Optimization (Trends and Future Directions)(ICRITO), 1-6.

Ali Alhosseini, S., Bin Tareaf, R., Najafi, P., & Meinel, C. (2019, May). Detect me if you can: Spam bot detection using inductive representation learning. In *Companion proceedings of the 2019 World Wide Web Conference* (pp. 148-153).

Alikhani, P., Brunner, N., Crépeau, C., Designolle, S., Houlmann, R., Shi, W., ... & Zbinden, H. (2021).

Experimental relativistic zero-knowledge proofs. *Nature, 599*(7883), 47-50.

Anoop, V. S., & Goldston, J. (2022). Decentralized finance to hybrid finance through blockchain: a case-study of Acala and Current. *Journal of Banking and Financial Technology, 6*(1), 109-115.

Barker, B., Souppaya, M., & Newhouse, W. (2021). Migration to post-quantum cryptography. National Institute of Standards and Technology and National Cybersecurity Center of Excellence. Retrieved from https://www.nccoe.nist.gov/sites/default/files/legacy-files/pqc-migration-project-description-final.pdf.

Béres, F., Seres, I. A., Benczúr, A. A., & Quintyne-Collins, M. (2021, August). Blockchain is watching you: Profiling and deanonymizing Ethereum users. In *2021 IEEE international conference on decentralized applications and infrastructures (DAPPS),* 69-78.

Berlee, A. (2017). Digital inheritance in the Netherlands. *J. Eur. Consumer & Mkt. L., 6*, 256-260.

Brown, J. H. (2019). Online tools under RUFADAA: The next evolution in estate planning or a flash in the pan. *Prob. & Prop., 33*, 60-63.

Browne, D., & Darcy, N. (2020). Wills and estates: Digital records and estate planning. *LSJ: Law Society Journal*, (66), 82-83.

Burdges, J., Cevallos, A., Czaban, P., Habermeier, R., Hosseini, S., Lama, F., ... & Wood, G. (2020). Overview of polkadot and its design considerations. *arXiv preprint arXiv:2005.13456*.

Bushmelev, F., Khlobystova, A., Abramov, M., & Livshits, L. (2022, November). Deep machine learning techniques in the problem of estimating the expression of psychological characteristics of a social media user. In *International Conference Artificial Intelligence in Engineering and Science* (pp. 315-324). Cham: Springer International Publishing.

Buterin, V. (2021a). Soulbound. Retrieved from: https://vitalik.ca/general/2022/01/26/soulbound.html

Buterin, V. (2021b). Why we need wide adoption of social recovery wallets. Retrieved from: https://vitalik.ca/general/2021/01/11/recovery.html

Candan, A., & Yaşlak, H. I. (2021). Fikhin kripto paralara bakişi: Bitcoin örneği. *Van İlahiyat Dergisi, 9*(15), 1-39.

Chaffer, T. J., & Goldston, J. (2022). On the existential basis of self-sovereign identity and soulbound tokens: An examination of the "self" in the age of web3. *Journal of Strategic Innovation and Sustainability, 17*(3), 1-9.

Chervinski, J. O., Yu, J., & Xu, X. (2021). Characterizing blockchain interoperability systems from an architecture perspective. In *International Summit Smart City 360°* (pp. 504-520). Cham: Springer International Publishing.

Chun, K. N., Park, S. Y., & Kim, M. S. (2022). Understanding decentralized autonomous organizations (DAOs) as a reaction to corporate governance problems. *Smatoos Business Review*, 33717.

Clinton, L. (2023). Fixing American cybersecurity: Creating a strategic public-private partnership. Georgetown University Press.

Cremation Institute. (2020, July). Cryptocurrency estate planning study. Retrieved from: https://cremationinstitute.com/crypto-estate-planning-study/

Diamantopoulou, V., Androutsopoulou, A., Gritzalis, S., & Charalabidis, Y. (2018, May). An assessment of privacy preservation in crowdsourcing approaches: Towards GDPR compliance. In *2018 12th International Conference on Research Challenges in Information Science (RCIS)* (pp. 1-9). IEEE.

Digital Assets Daily. (2022). Interview with Tyler Doussan from Safe Haven (SHA). Retrieved from:

https://www.youtube.com/watch?v=RrBESAKm0Hk.

Emmert, F. (2022). Cryptocurrencies: The impossible domestic law regime?. *The American Journal of Comparative Law, 70*(Supplement_1), i185-i219.

Ferdous, M. S., Chowdhury, F., & Alassafi, M. O. (2019). In search of self-sovereign identity leveraging blockchain technology. *IEEE access*, *7*, 103059-103079.

Flunger, R., Mladenow, A., & Strauss, C. (2022). Game analytics—business impact, methods and tools. *Developments in Information & Knowledge Management for Business Applications: Volume 3*, 601-617.

Gross, C., Lorek, K., & Richter, F. (2017). Attitudes towards inheritance taxation–results from a survey experiment. *The Journal of Economic Inequality, 15*, 93-112.

Guidi, B. (2020). When blockchain meets online social networks. *Pervasive and Mobile Computing, 62*, 101131.

Haley, T. D. (2023). Embracing digital. North Carolina Law Review, 101(3), 619-676.

Hewa, T., Ylianttila, M., & Liyanage, M. (2021). Survey on blockchain based smart contracts: Applications, opportunities and challenges. *Journal of network and computer applications, 177*, 102857.

Hirsch, A. J. (2020). Technology adrift: In search of a role for electronic wills. *BCL Rev., 61*, 828-903.

Jia, R., & Yin, S. (2022, November). To EVM or not to EVM: Blockchain compatibility and network effects. In *Proceedings of the 2022 ACM CCS Workshop on Decentralized Finance and Security,* 23-29.

Johnson, D. (2023). The more the merrier? Issues arising from co-trustees administering trusts. *Estate Planning & Community Property Law Journal, 15*(1), 36-90.

Johnson, S., Robinson, P., & Brainard, J. (2019). Sidechains and interoperability. *arXiv preprint arXiv:1903.04077*.

Kansal, M., Singh, A. K., & Dutta, R. (2022). Efficient multi-signature scheme using lattice. *The Computer Journal, 65*(9), 2421-2429.

Khanum, S., & Mustafa, K. (2023). A systematic literature review on sensitive data protection in blockchain applications. *Concurrency and Computation: Practice and Experience, 35*(1), e7422.

Kharitonova, J. S. (2021). Digital assets and digital inheritance. *Law & Digital Technology, 1*(4), 19-26.

Komarevceva, I., Melnichuk, M., Zhuravleva, O., Shulzhenko, I., & Zakharova, G. (2022). Development trends of inheritance law in the context of the digital economy. In Business 4.0 as a Subject of the Digital Economy (pp. 953-957). Springer, Cham.

Kopf, S. (2020). "Rewarding good creators": corporate social media discourse on monetization schemes for content creators. *Social Media+ Society*, *6*(4), 2056305120969877.

Korol, M. (2022). Social media and the Stored Communications Act: Translating the statutory bar on disclosure of private communications from civil to criminal discovery. *Loy. LAL Rev., 55*, 927-957.

Kurniawan, I. G. A., Lulo, L. D. D. M., & Disantara, F. P. (2023). IUS Constituendum of expert advisor in commodity futures trading: A legal certainty. *Jurnal IUS Kajian Hukum dan Keadilan, 11*(1), 31-45.

Lee, S., Kirk, A., Kirk, E. A., Karunanayake, C., O'Connell, M. E., & Morgan, D. (2019). Factors associated with having a will, power of attorney, and advanced healthcare directive in patients presenting to a rural and remote memory clinic. *Canadian Journal of Neurological Sciences, 46*(3), 319-330.

Lee, Y. K., & Jeong, J. (2021). Securing biometric authentication system using blockchain. *ICT Express, 7*(3), 322-326.

Lucsok, P. (2020, April). Social recovery on Substrate. Parity Technologies. Retrieved December 31, 2022, from https://www.parity.io/blog/social-recovery-on-substrate/.

Manolache, M. A., Manolache, S., & Tapus, N. (2022). Decision making using the blockchain proof of authority consensus. *Procedia Computer Science*, *199*, 580-588.

Ngo, T. T., Dang, T. A., Huynh, V. V., & Le, T. C. (2023). A systematic literature mapping on using blockchain technology in identity management. *IEEE Access, 11*, 26004-26032.

Njui, J. P. (2018, April). $500 million in Ripple (XRP) possibly lost forever after billionaire owner dies unexpectedly. *Ethereum World News*. Retrieved from https://ethereumworldnews.com/500-million-in-ripple-xrp-possibly-lost-forever-after-billionaire-owner-dies-unexpectedly/

Oli. (2022). DeQuest goes into partnership with RMRK. Retrieved from: https://medium.com/dequest/dequest-goes-into-partnership-with-rmrk-6746a4ac8108.

Orabi, M., Mouheb, D., Al Aghbari, Z., & Kamel, I. (2020). Detection of bots in social media: a systematic review. *Information Processing & Management, 57*(4), 102250.

Panda, S. K., Sathya, A. R., & Das, S. (2023). Bitcoin: Beginning of the cryptocurrency era. In *Recent Advances in Blockchain Technology: Real-World Applications* (pp. 25-58). Cham: Springer International Publishing.

Petrowski, J., & Tabrizi, S. (2020). Substrate Social Recovery Pallet Code Walkthrough with Joe Petrowski and Shawn Tabrizi. Retrieved from https://youtu.be/rRnWKDaTb9E.

Qi, J. (2023). The terror management function of descendent continuity: Evidence that descendent continuity acts as a distal and proximal defense. *Personality and Social Psychology Bulletin, 49*(1), 125-137.

Rahardja, U., Hidayanto, A. N., Putra, P. O. H., & Hardini, M. (2021). Immutable ubiquitous digital certificate authentication using blockchain protocol. *Journal of Applied Research and Technology, 19*(4), 308-321.

Saidakhrarovich, G. S. (2022). Digitalization in inheritance law. *World Bulletin of Management and Law, 10*, 18-30.

Satybaldy, A., Nowostawski, M., & Ellingsen, J. (2020). Self-sovereign identity systems: Evaluation framework. *Privacy and Identity Management. Data for Better Living: AI and Privacy: 14th IFIP WG 9.2, 9.6/11.7, 11.6/SIG 9.2. 2 International Summer School, Windisch, Switzerland, August 19–23, 2019, Revised Selected Papers 14*, 447-461.

Schouppe, J. (2022). *U.S. Patent No. 11,316,668*. Washington, DC: U.S. Patent and Trademark Office.

Scott, I. J., de Castro Neto, M., & Pinheiro, F. L. (2023). Bringing trust and transparency to the opaque world of waste management with blockchain: A Polkadot parathread application. *Computers & Industrial Engineering*, 109347.

Sehati, I. N. (2021). Beyond the grave: A fiduciary's access to a decedent's digital assets. *Cardozo L. Rev., 43*, 745-782.

Shaik, C. (2020). Securing cryptocurrency wallet seed phrase digitally with blind key encryption. *International Journal on Cryptography and Information Security (IJCIS), 10*(4).

Sheridan, P. (2020). Inheriting digital assets: does the revised uniform fiduciary access to digital assets act fall short?. *Ohio St. Tech. LJ, 16*, 363-394.

Shrestha, P., & Bhattarai, P. C. (2021). Application of case study methodology in the exploration of inclusion in education. *American Journal of Qualitative Research*, *6*(1), 73-84.

Shuaib, M., Daud, S. M., & Alam, S. (2021). Self-sovereign Identity framework development in compliance with Self sovereign Identity principles using components. *Int. J. Mod. Agric, 10*(2), 3277-3296.

Singh, R. G., Shrivastava, A., & Ruj, S. (2022). A digital asset inheritance model to convey online persona posthumously. *International Journal of Information Security*, *21*(5), 983-1003.

Skibinsky, M., Dodis, Y., Spies, T., & Ahmad, W. (2018). Decentralized storage of crypto assets via hierarchical Shamir's secret sharing. *Vault12 Whitepaper*.

Smyth, C., & Peiros, K. (2017). An international will or a will in each jurisdiction?. *LSJ: Law Society of NSW Journal*, (34), 78-79.

Soltani, R., Nguyen, U. T., & An, A. (2021). A survey of self-sovereign identity ecosystem. *Security and Communication Networks*, 2021, 1-26.

Sun, N., Zhang, Y., & Liu, Y. (2022). A privacy-preserving kyc-compliant identity scheme for accounts on all public blockchains. *Sustainability, 14*(21), 14584.

Tabrizi, S. (2020) Substrate Recovery Pallet - Shawn Tabrizi. Retrieved from https://youtu.be/ZfhEAzRCFBc.

Tatar, S., & Akmaljon, A. (2023, February). Digitalization in inheritance law. Horizon: Journal of Humanity and Artificial Intelligence, 2(2), 15-22.

Teisserenc, B., & Sepasgozar, S. M. (2022). Software architecture and non-fungible tokens for digital twin decentralized applications in the built environment. *Buildings, 12*(9), 1447.

Thomas, A. M., Ramaguru, R., & Sethumadhavan, M. (2022). Distributed identity and verifiable claims using Ethereum standards. In *Inventive Communication and Computational Technologies: Proceedings of ICICCT 2021,* 621-636.

Tsepeleva, R., & Korkhov, V. (2022). Building DeFi applications using cross-blockchain interaction on the wish swap platform. *Computers, 11*(6), 99.

Usman, A. B., Cole, N., Asplund, M., Boeira, F., & Vestlund, C. (2023, April). Remote attestation assurance arguments for trusted execution environments. In Proceedings of the 2023 ACM Workshop on Secure and Trustworthy Cyber-Physical Systems (pp. 33-42).

Valeonti, F., Bikakis, A., Terras, M., Speed, C., Hudson-Smith, A., & Chalkias, K. (2021). Crypto collectibles, museum funding and OpenGLAM: challenges, opportunities and the potential of non-fungible tokens (NFTs). *Applied Sciences, 11*(21), 9931.

van Rijmenam, M. (2022). *Step into the metaverse: How the immersive internet will unlock a trillion-dollar social economy*. John Wiley & Sons.

Waller, J. (2022, October). Phat contract examples. Retrieved from https://github.com/HashWarlock/phat-contract-examples/tree/master/examples/phat-s3.

Weller, K., & Kinder-Kurlanda, K. E. (2016, May). A manifesto for data sharing in social media research. In *Proceedings of the 8th ACM Conference on Web Science* (pp. 166-172).

Weyl, E. G., Ohlhaver, P., & Buterin, V. (2022). Decentralized society: Finding web3's soul. *Available at*

*SSRN 4105763*.

Wilson, R. (2022). AB 633: Protecting family property from unscrupulous developers and real estate speculators. *University of the Pacific Law Review, 53*(2), 272-286.

Wood, G. (2016). Polkadot: Vision for a heterogeneous multi-chain framework. *White paper, 21*(2327), 4662.

Wood, G. (2022). Substrate and Polkadot. Available at https://substrate.io/vision/substrate-and-polkadot/

Xue, Y., & Wang, J. (2022). Design of a blockchain-based traceability system with a privacy-preserving scheme of zero-knowledge proof. *Security and Communication Networks*, *2022*.

Zheng, Y. (2023). A Primitive Study on Privacy Protection of Digital Heritage--From the Perspective of Social media accounts. *Advances in Economics and Management Research*, *4*(1), 16-16.

Zhuang, Y., Shyu, C. R., Hong, S., Li, P., & Zhang, L. (2023). Self-sovereign identity empowered non-fungible patient tokenization for health information exchange using blockchain technology. *Computers in Biology and Medicine, 157*, 106778.

Zulkiffli, S. N. H., Zawawi, M. N. A., & Rahim, F. A. (2020, August). Passive and active reconnaissance: a social engineering case study. In *2020 8th International Conference on Information Technology and Multimedia (ICIMU)* (pp. 138-143). IEEE.